%% file: Hayashi.tex
\documentclass[graybox]{svmult}

\usepackage{type1cm}        % activate if the above 3 fonts are
\usepackage{makeidx}         % allows index generation
\usepackage{graphicx}        % standard LaTeX graphics tool
\usepackage{multicol}        % used for the two-column index
\usepackage[bottom]{footmisc}% places footnotes at page bottom

\usepackage{newtxtext}       % 
\usepackage[varvw]{newtxmath}       % selects Times Roman as basic font

\def\U{\mathop{\rm U}}
\def\ce{\mathop{\rm ce}}
\def\cH{\mathcal{H}}
\def\cV{\mathcal{V}}
\def\sff{\mathsf{f}}
\def\Tr{\mathop{\rm Tr}}
\def\bZ{\mathbb{Z}}
\def\bR{\mathbb{R}}
\def\argmax{\mathop{\rm argmax}}

\makeindex             % used for the subject index
\begin{document}

\title*{Special functions in quantum phase estimation}
\author{Masahito Hayashi}
\institute{Masahito Hayashi\at 
Shenzhen Institute for Quantum Science and Engineering, Southern University of Science and Technology, Shenzhen,518055, China,
International Quantum Academy (SIQA), Futian District, Shenzhen 518048, China,
Graduate School of Mathematics, Nagoya University, Nagoya, 464-8602, Japan.
\email{hayashi@sustech.edu.cn, masahito@math.nagoya-u.ac.jp}
}
\maketitle

\abstract{
This paper explains existing results for the application of special functions to phase estimation, which is 
a fundamental topic in quantum information.
We focus on two special functions. 
One is prolate spheroidal wave function, which approximately gives 
the maximum probability that the difference between the true parameter and the estimate is smaller than a certain threshold.
The other is Mathieu function, which exactly gives the optimum estimation under the energy constraint.
It also characterizes the uncertainty relation for 
the position and the momentum for periodic functions.}

\section{Introduction}
\label{sec:2}
It is well known that quantum system has group symmetry.
Therefore, various quantum information processing can utilize
group symmetry to enhance or optimize various operations.
One typical example is the estimation of the unknown unitary operation.
In this problem setting, the set of possible unitary operations often 
forms a group representation.
When the input state is fixed to a certain state,
this problem can be considered as a special case of 
the estimation of the unknown state under the group symmetric model.
For this type of state estimation, 
Holevo formulated a systematic group symmetric approach \cite{Holevo,Holevo2}.
Holevo's approach 
is known as a powerful tool for state estimation \cite{H98,Group2}.
By using Holevo's approach, 
the above estimation problem of the unknown unitary operation
has been formulated in a general form by \cite{CDS,CMP}.

The simplest case of the estimation of the unknown unitary operation
is phase estimation, which is formulated as optimizations of 
estimating methods of an unknown element of 
$\U(1)$.
A group symmetric approach works well for this problem.
Interestingly, 
although this problem can be formulated dependently 
of the choices of error function and available input systems or input states,
the optimal solution under several special cases
can be characterized by special functions.
This paper surveys existing results for these relations between 
special functions and 
the optimal solution under several examples of 
phase estimation.
In particular, this paper focuses on two special functions,
prolate spheroidal wave function and 
Mathieu function.

Prolate spheroidal wave function approximately gives 
an optimal input state 
to maximize the probability that the difference between the true parameter and the estimate is smaller than a certain threshold.
Mathieu function gives 
an optimal input state under a certain energy constraint.
This characterization can be used for the uncertainty relation between 
the position and the momentum on the periodic function space.
In this way, these two special functions play a central role in phase estimation.

The remaining of this paper is organized as follows.
Section 2 gives the formulation of phase estimation.
Section 3 discusses the phase estimation under more specific examples.
This section presents the relation between  
Prolate spheroidal wave function and phase estimation.
Section 4 addresses the phase estimation under the energy constraint.
This section presents the relation between  
Mathieu function and phase estimation.
Section 5 applies the result in Section 4 to 
the uncertainty relation between 
the position and the momentum on the periodic function space.

\begin{figure}[b]
\sidecaption
% For example, with the graphicx style use
\includegraphics[scale=.45]{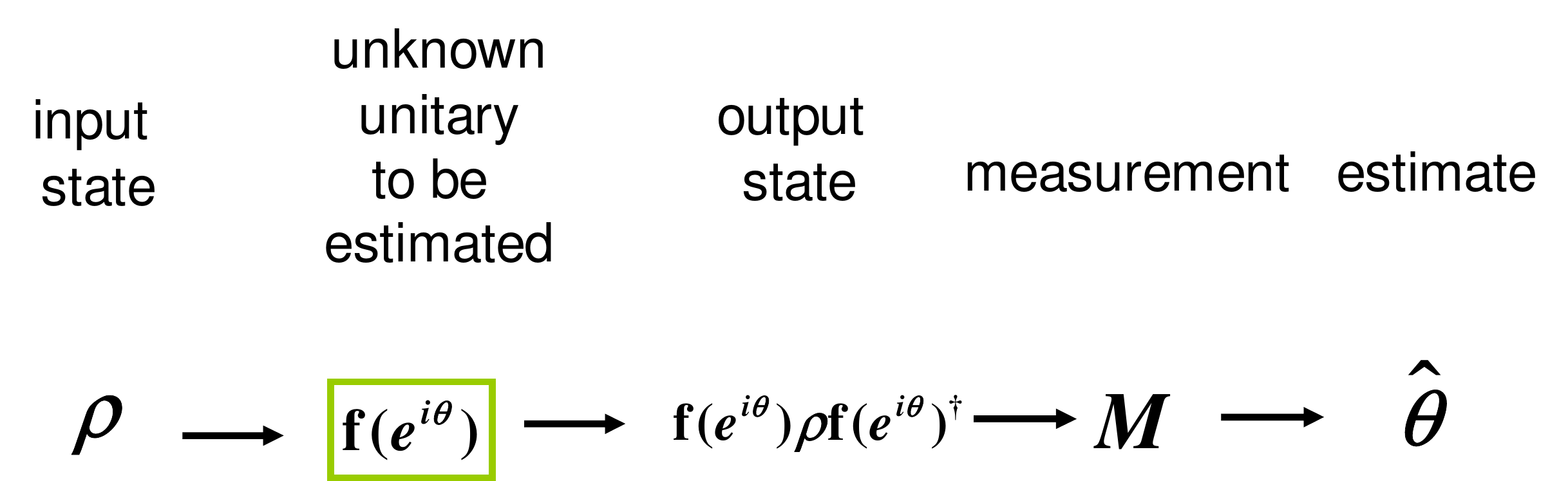}
%
% If no graphics program available, insert a blank space i.e. use
%\picplace{5cm}{2cm} % Give the correct figure height and width in cm
%
\caption{This figure expresses the process to estimate the action of the unknown element of $\U(1)$.}
\label{fig:1}       % Give a unique label
\end{figure}

\section{Formulation}
\label{sec:1}
We estimate the unknown application of an element of $\U(1)$ in various settings.
To cover various settings, 
this problem is formulated as follows.
First, we consider a fixed unitary representation $\sff$ of the group $\U(1)$ on a Hilbert space $\cH$, which 
represents our physical system.
We are allowed to choose the input state $\rho$
and the quantum measurement on the system $\cH$ to get our estimate in $\U(1)$.
The quantum measurement on the system $\cH$ is rewritten as a positive operator-valued measure on $\cH$,
which is given as ${\cal M}:=(M_\theta)_{\theta \in [0,2\pi)}$ with the condition
\begin{align}
\int_0^{2\pi }M_\theta d\theta=I\label{HA1}
\end{align}
by identifying $\U(1)$ with $[0,2\pi)$.
Our estimation scheme for the unknown application $\sff(e^{i \theta})$
with $e^{i \theta}\in \U(1) $ is formulated as Fig. \ref{fig:1} 
\cite{CMP,LP,BDM,PLA,IH09}.

When the true unitary action is $\sff(e^{i\theta})$,
the output $\hat{\theta} \in [0,2\pi)$ is generated by the distribution
$\Tr \sff(e^{i\theta}) \rho \sff(e^{i\theta})^\dagger M_{\hat{\theta}}d \hat{\theta}$.
To evaluate the precision of our estimate, 
we consider the error function $R(\theta,\hat{\theta})$.
For the symmetry of our problem setting, we impose 
the symmetric condition 
\begin{align}
R(\theta,\hat{\theta})=R(0,\hat{\theta}-\theta)=R(0,\hat{\theta}-\theta+2n\pi)
\label{HA2}
\end{align}
 with any integer $n$. 
Then, the average error is calculated as a function of 
$\theta,\rho,{\cal M}$ \cite{CMP};
\begin{align}
{\cal R}[\sff,R,\theta,\rho,{\cal M}]:=
\int_0^{2\pi}R(\theta,\hat{\theta})
\Tr \sff(e^{i\theta}) \rho \sff(e^{i\theta})^\dagger M_{\hat{\theta}}d \hat{\theta}.
\end{align}
It is natural to focus on 
the worst value
${\cal R}_{\max}[\sff,R,\rho,{\cal M}]:=
\max_{\theta}{\cal R}[\sff,R,\theta,\rho,{\cal M}]$
or 
the average value
${\cal R}_{av}[\sff,R,\rho,{\cal M}]:=
\frac{1}{2\pi}\int_0^{2\pi}{\cal R}[\sff,R,\theta,\rho,{\cal M}] d\theta$
with respect to the unknown parameter $\theta$ \cite{CMP}.
We consider the following minimizations \cite{CMP}
\begin{align}
{\cal R}_{\max}[\sff,R]:=
\min_{\rho,{\cal M}}
{\cal R}_{\max}[\sff,R,\rho,{\cal M}],\quad
{\cal R}_{av}[\sff,R]:=
\min_{\rho,{\cal M}}
{\cal R}_{av}[\sff,R,\rho,{\cal M}].
\end{align}

\begin{figure}[b]
\sidecaption
% For example, with the graphicx style use
\includegraphics[scale=.45]{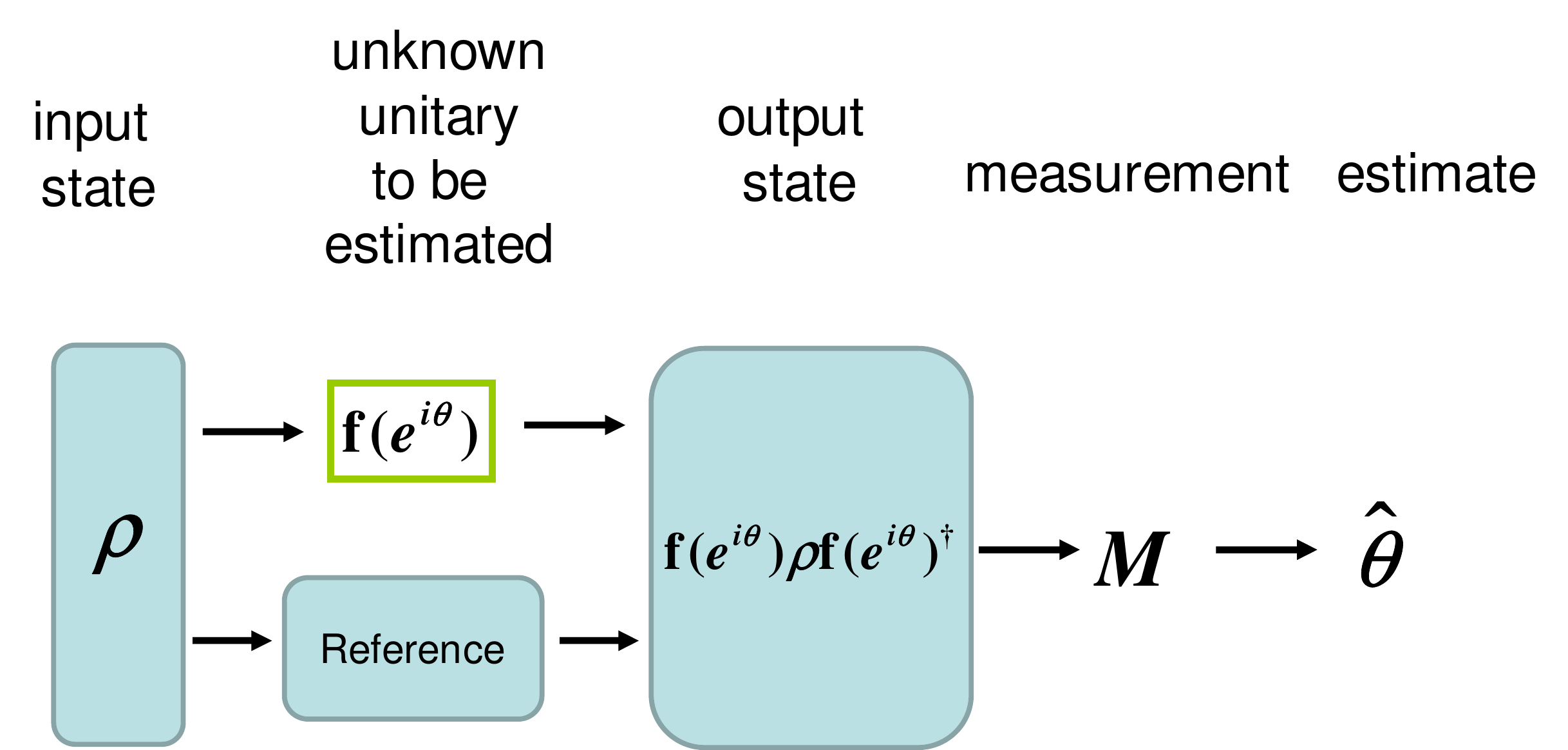}
%
% If no graphics program available, insert a blank space i.e. use
%\picplace{5cm}{2cm} % Give the correct figure height and width in cm
%
\caption{This figure expresses the process to estimate the action of the unknown element of $\U(1)$ with the reference system.
The input system might be an entangled state between the system $\cH$ and the reference system $\cH_R$.
}
\label{fig:2}       % Give a unique label
\end{figure}

To discuss the above problems, 
we consider a detailed structure.
An irreducible representation of $\U(1)$ 
is characterized by an integer $n \in \bZ$ and 
has a one-dimensional representation space $\cH_n$.
This representation is denoted as 
$\sff_n$ and is defined as $\sff_{n}(e^{i \theta})=e^{i n\theta} $.

Now, we consider a general representation $\sff$ of $\U(1)$ and 
its representation space $\cH$.
Let $S$ be the set of indexes $n$ whose corresponding irreducible representation
$\sff_n$ is contained in $\sff$.
We denote the multiplicity of $\sff_n$ in $\sff$ by $m_n$,
and define an $m_n$-dimensional space by $\cV_n$.
Then, the representation space $\cH$ is written as
$\oplus_{n \in S} \cH_n \otimes \cV_n$,
where the group $\U(1)$ acts only on $\cH_n$.
That is, for $x=\oplus_{n \in S} x_n \otimes v_n\in \oplus_{n \in S} \cH_n \otimes \cV_n$,
we have
\begin{align}
\sff(g)x =\bigoplus_{n \in S} (\sff_n(g)x_n) \otimes v_n
\end{align}
for $g \in \U(1)$.

This formulation contains the case when 
the input state is an entangled state between the system $\cH$ and a reference system
$\cH_R$ as Fig. \ref{fig:2}
because the joint system $\cH\otimes \cH_R$ has the form 
$\oplus_{n \in S} \cH_n \otimes \cV_n$.

When the multiplicity $m_n$ is one for any $n \in S$, 
the representation $\sff$ is called multiplicity-free 
with $S$ and is denoted by $\sff_S$.
Under the representation $\sff_S$, we denote a normalized vector in $\cH_n$ by $e_n$.
The representation space of the representation $\sff_S$ is 
the space $\cH_S$ spanned by the orthogonal
vectors $\{e_n\}_{n \in S}$.
Under the representation $\sff_S$, we consider the following types of 
positive operator-valued measure.
Consider a vector $|w\rangle:= \sum_{n\in S}|e_n\rangle $. We choose $M_{\hat{\theta}}:=
\frac{1}{2\pi} \sff_S(e^{i\hat{\theta}})^\dagger
|w \rangle \langle w|\sff_S(e^{i\hat{\theta}})$, which satisfies the condition \eqref{HA1}
for POVM.
This POVM is written as ${\cal M}_w$.
Also, an element $|\phi\rangle$ of the vector space $\cH_n$
can be identified with $(\phi_n)_{n \in S}$ through the relation 
$|\phi\rangle= \sum_{n \in S} \phi_n|e_n\rangle $.
We define the Fourier transform $ {\cal F}[\phi](\hat{\theta})$ as
$\sum_{n \in S} \phi_n e^{i n \hat{\theta}}
=\langle w|\sff_S(e^{i\hat{\theta}}) |\phi\rangle$.
Then, as shown in \cite[Lemma 1 and Theorem 1]{CMP} \cite{CDS}, we have
\begin{align}
{\cal R}_{\max}[\sff,R]=&
{\cal R}_{av}[\sff,R]
=
\min_{|\phi\rangle \in \cH_S }
{\cal R}[\sff_S,R,0,|\phi\rangle\langle\phi|,{\cal M}_w]\nonumber \\
=&
\min_{|\phi\rangle \in \cH_S }
\frac{1}{2\pi}\int_{0}^{2\pi} R(0,\hat{\theta})
 \langle w|\sff_S(e^{i\hat{\theta}}) |\phi\rangle \langle \phi|
\sff_S(e^{i\hat{\theta}})^\dagger |w \rangle
d \hat{\theta}\nonumber \\
=&
\min_{|\phi\rangle \in \cH_S }
\frac{1}{2\pi}\int_{0}^{2\pi} R(0,\hat{\theta})
|{\cal F}[\phi](\hat{\theta})|^2
d \hat{\theta}.\label{HA7}
\end{align}

\section{Constraint for available irreducible representation}
In this section, we consider several examples where
available irreducible representation is restricted.
We assume that $R$ is given as 
$R_{\sin}(\theta,\hat{\theta}):=2 \sin^2 \frac{\theta-\hat{\theta}}{2} 
=1-\cos(\theta-\hat{\theta})$.
We consider a typical representation $\sff_{\{0,1\}}$.
We often consider its $n$-fold tensor product representation $\sff_{\{0,1\}}^{\otimes n}$.
In this representation, the set of indexes $S$ is $\{0,1, \ldots, n\}$.
Hence, it is sufficient to address $\sff_{\{0,1, \ldots, n\}}$.
Then, the minimization \eqref{HA7} is calculated as
\begin{align}
{\cal R}_{\max}[\sff_{\{0,1\}}^{\otimes n},R_{\sin}]=&
{\cal R}_{av}[\sff_{\{0,1\}}^{\otimes n},R_{\sin}]
=
{\cal R}_{av}[\sff_{\{0,1,\ldots,n\}},R_{\sin}]
\nonumber \\
=&
\min_{|\phi\rangle \in \cH_{\{0,1, \ldots, n\}} }
\frac{1}{2\pi}\int_{0}^{2\pi} 
2 \sin^2 \frac{\hat{\theta}}{2} 
|{\cal F}[\phi](\hat{\theta})|^2
d \hat{\theta} \nonumber \\
=&
\min_{|\phi\rangle \in \cH_{\{0,1, \ldots, n\}} }
1-\frac{1}{2}
\sum_{j=0}^{n-1} 
(\overline{\phi}_j \phi_{j+1}
+\phi_j \overline{\phi}_{j+1}).
\end{align}
For the derivation of the final step, see 
\cite{Holevo,Holevo2}
\cite{BDM}, \cite[Section 2]{PLA} \cite[Theorem 7]{CMP}.

In fact, the maximum eigenvalue of the operator 
$\frac{1}{2}\sum_{j=0}^n (|e_j\rangle \langle e_{j+1}|+|e_{j+1}\rangle \langle e_j|)$
is $\cos \frac{\pi}{n+1}$, and its corresponding eigenvector is 
$C \sum_{j=0}^n \sin\frac{j\pi}{n+1}|e_j\rangle$ with a normalizing constant $C$ \cite[Theorem 7]{CMP}.
Hence, 
the above minimum is 
\begin{align}
1-\cos \frac{\pi}{n+1}=
2 \sin^2 \frac{\pi}{2(n+1)}, \label{HA10}
\end{align}
which asymptotically behaves as 
$\frac{\pi^2}{2 n^2} $.
This type of analysis was extended to the case with the group SU(2) \cite{BBM,CDPS2,PLA}.
In this case, the error is inverse proportional to $n^2$.
This scaling is called Heisenberg scaling.

\begin{remark}
Here, it is better to remark that many papers
discussed Heisenberg scaling in a misleading way \cite{GLM,GLM2,NOOST,OHNOST,JKFABBM}. 
The above discussion calculated the minimum error.
To discuss the asymptotic behavior of the minimum error,  
instead of the above calculation,
these papers employ the relation between the estimation error and Fisher information.
The estimation error is lower bounded by the inverse of Fisher information.
The attainability of this lower bound is not trivial in general.
For example, 
In the case of state estimation, 
this lower bound can be attained by a two-step method under a natural regularity condition \cite{HM}.
However, in the case of unitary estimation,
this lower bound cannot be attained.
In particular, the lower bound given by the maximum Fisher information
is strictly smaller than the optimal minimum estimation error even in the level of the first order coefficient \cite{CMP2}.
These papers considered that 
the maximum Fisher information gives the estimation error even in this case
while
Fisher information approach does not work for the Heisenberg scaling of the estimation error in phase estimation.
\end{remark}

Next, we discuss the asymptotic behavior in another way \cite[Section 4]{IH09}.
For simple analysis, we focus on the representation 
$\sff_{\{-N, \ldots, N\}}$ instead of $\sff_{\{0,1, \ldots, n\}}$.
We consider the function space $L^2([-1,1])$ and its dense subset 
$L_c^2([-1,1]):= L^2([-1,1]) \cap C([-1,1])$, where 
$L^2([-1,1])$ is the set of square integrable functions on $[-1,1]$ and 
$C([-1,1])$ is the set of continuous functions on $[-1,1]$.
Given a normalized continuous function $\psi\in L_c^2([-1,1])$,
we choose $\phi^{(n)} \in  \cH_{\{-N, \ldots, N\}}$ as
the normalized vector of $(\psi(\frac{j}{N}))_{j=-N}^N$.
We define the Fourier transform ${\cal F}$ on $L^2(\bR) $ as
\begin{align}
{\cal F}[\psi](t):= \frac{1}{\sqrt{2\pi}}\int_{-\infty}^\infty e^{i t x}\psi(x) dx.
\end{align}
Then, using $t=N \hat{\theta}$,
we have
\begin{align}
&\frac{N^2}{2\pi}\int_{0}^{2\pi} 
2 \sin^2 \frac{\hat{\theta}}{2} 
|{\cal F}[\phi](\hat{\theta})|^2
d \hat{\theta} \nonumber\\
\cong &
\frac{1}{2}
\int_{-\infty}^\infty t^2 | {\cal F}[\psi](t)|dt
= \frac{1}{2}
\langle {\cal F}[\psi]|Q^2|{\cal F}[\psi]\rangle
= \frac{1}{2}
\langle \psi|P^2|\psi\rangle.\label{HA13}
\end{align}
Here $Q$ is the multiplication operator and $P$ is the momentum operator defined as
$P \psi (x):= i \frac{d}{dx}\psi (x)$.
In fact, the minimum eigenvalue of $P^2$ on the function space $L^2([-1,1])$ 
is $\frac{\pi^2}{4}$.
Hence, the minimum of \eqref{HA13} is 
$\frac{\pi^2}{8}$, which coincides the asymptotic behavior of \eqref{HA10} with $n=2N$.

Next, given an real number $T>0$, we maximize the probability 
satisfying the condition 
$-\frac{T}{N} < \hat{\theta}-\theta <\frac{T}{N}$ \cite[Section 5]{IH09}.
For this aim, we choose the function $R(\theta,\hat{\theta})$ as the probability 
satisfying the condition $ |\hat{\theta}-\theta| \ge \frac{T}{N}$, which is denoted by
$R[T](\theta,\hat{\theta}) $.
Then, we have
\begin{align}
{\cal R}_{\max}[\sff_{\{-N,\ldots,N\}},R[T]]=&
{\cal R}_{av}[\sff_{\{-N,\ldots,N\}},R[T]]
=
{\cal R}_{av}[\sff_{\{-N,\ldots,N\}},R[T]]
\nonumber\\
=&
\min_{|\phi\rangle \in \cH_{\{-N, \ldots, N\}} }
1-\frac{1}{2\pi}\int_{-\frac{T}{N}}^{\frac{T}{N}} 
|{\cal F}[\phi](\hat{\theta})|^2
d \hat{\theta} .\label{HA16}
\end{align}

For simple analysis, we focus on the case when 
the vector $|\phi\rangle \in \cH_{\{-N, \ldots, N\}}$ is given in the above way.
As shown in \cite[Section 5]{IH09}, we have
\begin{align}
\frac{1}{2\pi}\int_{-\frac{T}{N}}^{\frac{T}{N}} 
|{\cal F}[\phi](\hat{\theta})|^2
d \hat{\theta} 
\cong
\int_{-T}^T | {\cal F}[\psi](t)|dt.
\end{align}
We define the projection $\Pi_T$ corresponding to the event that the spectral of $Q$ belongs to $[-T,T]$.
Since $\psi\in L_c^2([-1,1])$ belongs to the range of the projection $\Pi_1$,
we have
\begin{align}
\int_{-T}^T | {\cal F}[\psi](t)|dt
= \langle {\cal F}[\psi]| \Pi_T|{\cal F}[\psi]\rangle
= \langle \psi| {\cal F}\Pi_T {\cal F}|\psi\rangle
= \langle \psi| \Pi_1{\cal F}\Pi_T {\cal F}\Pi_1|\psi\rangle.
\end{align}
The problem \eqref{HA16} is converted to the maximization of 
$\langle \psi| \Pi_1{\cal F}\Pi_T {\cal F}\Pi_1|\psi\rangle$.
To discuss the maximum eigenvalue of the operator $\Pi_1{\cal F}\Pi_T {\cal F}\Pi_1$,
we consider the prolate spheroidal wave function $\psi_T$, which is the solution of the differential equation
\begin{align}
\frac{d}{dx}(1-x^2)\frac{d\psi}{dx}+(\xi(T)-T^2x^2)\psi(x)=0,
\end{align}
where $\xi(T)$ is a real number depending on $T$\footnote{For the relation between $\xi(T)$ 
and $T$, see Slepian and Pollak \cite{SP}.}.
Slepian and Pollak \cite{SP} showed that the function $\psi_T$ is the eigenfunction of 
the operator $\Pi_1{\cal F}\Pi_T {\cal F}\Pi_1$ with the 
maximum eigenvalue $\lambda(T)$, which behaves as \cite{Slepian}
\begin{align}
1-\lambda(T)\cong 4 \sqrt{\pi T}e^{-2T} \Big(1-\frac{3}{32T}+O(T^2)\Big).
\end{align}
In this way, the asymptotic bahavior of the problem \eqref{HA16}
is closely linked to a special function, the prolate spheroidal wave function.

\section{Energy constraint}
Now, we impose an energy constraint on the input state on $\cH$ for a representation $\sff$ \cite[Section 11]{CMP}.
We define the Hamiltonian $H$ on $\cH$ as
\begin{align}
H:= \sum_{j \in S}j^2 I_j, 
\end{align}
where $I_j$ is the projection to the subspace $\cH_j \otimes \cV_j $.
Then, we impose the following energy constraint to the input state $\rho$ as
\begin{align}
\Tr \rho H \le E.\label{HA23}
\end{align}
In the following, we consider the case with $S=\bZ$, and 
denote the set of states with the condition \eqref{HA23} by ${\cal S}_E$.
We consider the following minimizations
\begin{align}
{\cal R}_{\max}[\sff,R,E]:=&
\min_{\rho \in {\cal S}_E,{\cal M}}
{\cal R}_{\max}[\sff,R,\rho,{\cal M}],\\
{\cal R}_{av}[\sff,R,E]:=&
\min_{\rho\in {\cal S}_E,{\cal M}}
{\cal R}_{av}[\sff,R,\rho,{\cal M}].
\end{align}
Let $\cH_{\bZ,E}$ be the set of normalized vectors 
$\phi \in \cH_{\bZ}$ to satisfy the condition 
$\langle \phi |H|\phi\rangle \le E$.
When the error function $R$ satisfies the symmetric condition \eqref{HA2},
as shown in \cite[Theorem 2]{CMP} as a variant of \eqref{HA7},
we have
\begin{align}
{\cal R}_{\max}[\sff,R,E]=&
{\cal R}_{av}[\sff,R,E]
=
\min_{|\phi\rangle \in \cH_S }
{\cal R}[\sff_S,R,0,|\phi\rangle\langle\phi|,{\cal M}_w]\nonumber\\
=&
\min_{|\phi\rangle \in \cH_{\bZ,E} }
\frac{1}{2\pi}\int_{0}^{2\pi} R(0,\hat{\theta})
 \langle w|\sff_S(e^{i\hat{\theta}}) |\phi\rangle \langle \phi|
\sff_S(e^{i\hat{\theta}})^\dagger |w \rangle
d \hat{\theta}\nonumber\\
=&
\min_{|\phi\rangle \in \cH_{\bZ,E} }
\frac{1}{2\pi}\int_{0}^{2\pi} R(0,\hat{\theta})
|{\cal F}[\phi](\hat{\theta})|^2
d \hat{\theta}.\label{HA27}
\end{align}

To consider this problem, 
we define the function space 
$L^2_{p}((-\pi,\pi])$ as 
the space of the periodic square integrable functions with the period $2\pi$.
Then, we define the function space
$L^2_{p,even}((-\pi,\pi])$ as the space of even functions in
$L^2_{p}((-\pi,\pi])$.
Now, we choose $R_{\sin}(\theta,\hat{\theta})=2 \sin^2 \frac{\theta-\hat{\theta}}{2} 
=1-\cos(\theta-\hat{\theta})$.
Then, as shown in \cite[Theorem 6 and Eq. (97)]{CMP}, we have
\begin{align}
&\min_{|\phi\rangle \in \cH_{\bZ,E} }
\frac{1}{2\pi}\int_{0}^{2\pi} R_{\sin}(0,\hat{\theta})
|{\cal F}[\phi](\hat{\theta})|^2
d \hat{\theta}\nonumber\\
=&\kappa(E):= \min_{ \psi \in L^2_{p,even}(-\pi,\pi)}
\{ \langle \psi |I-\cos Q |\psi\rangle | 
\langle \psi |P^2 |\psi\rangle \le E, \|\psi \|=1\}\label{HA28}.
\end{align}

To calculate the function $\kappa$,
we define the function
\begin{align}
\gamma(s):
=&\min_{\psi \in L^2((-\pi,\pi]),\|\psi\|=1} 
\langle\psi| I-\cos Q +sP^2|\psi\rangle  \nonumber\\
=&\min_{\psi \in L^2((-\pi/2,\pi/2]),\|\psi\|=1} 
\langle\psi| I-\cos Q +sP^2|\psi\rangle .
\end{align}
Then, $\kappa(E)$ is given by the Legendre transform of $\gamma(s)$, i.e.,
as shown in \cite[Lemma 6]{CMP}, 
we have the formula 
\begin{align}
\kappa(E)= \max_{s >0}\gamma(s)-sE.\label{HA33}
\end{align}

The value $\gamma(s)$ can be characterized as the minimum value of $\gamma$ 
having the solution in
$L^2((-\pi/2,\pi/2])$ of the following differential equation.
\begin{align}
\frac{s}{4}\frac{d^2}{d\theta^2} \varphi(\theta)+(\gamma-1+\cos(2\theta))
\varphi(\theta)=0,
\end{align}
which is equivalent to
\begin{align}
\frac{d^2}{d\theta^2} \varphi(\theta)+( \frac{4(\gamma-1)}{s}+\frac{4}{s}\cos(2\theta))
\varphi(\theta)=0.
\end{align}

Now, we consider Mathieu equation:
\begin{align}
\frac{d^2}{d\theta^2}\varphi(\theta) +(a-2q \cos (2\theta))\varphi(\theta)=0.
\label{HA36}
\end{align}
A function $\varphi$ satisfies the above equation if and only if the function $\varphi$ 
is the eigenfunction of the differential operator $P^2 + 2q \cos(2Q)$. 
The operator $X(q) := P^2 + 2q \cos(2Q)$
preserves the subspace 
$L^2_{p,even}((-\frac{\pi}{2},\frac{\pi}{2}])$.
Then, we denote the minimum eigenvalue in 
$L^2_{p,even}((-\frac{\pi}{2},\frac{\pi}{2}])$ by $a_0(q)$, which is also 
the minimum eigenvalue in 
$L^2_{p}((-\frac{\pi}{2},\frac{\pi}{2}])$ \cite[Section 28.2]{Wolf}.
Mathieu function $\ce_0(\theta, q)$ is defined as 
the solution of \eqref{HA36} with $a_0(q)$ \cite[Section 28.2(vi)]{Wolf}.

Then, since $\gamma(s)$ is $\gamma$ in \eqref{HA36},
we have
\begin{align}
\gamma(s)= \frac{s a_0(\frac{2}{s})}{4}+1.
\end{align}
Hence, using the formula \eqref{HA33},
we have
\begin{align}
\kappa(E)=\max_{s>0} \frac{s a_0(\frac{2}{s})}{4}+1-sE.\label{E27}
\end{align}
The minimum in \eqref{HA28} is attained if and only if
$ {\cal F}[\psi](\theta)= \ce_0(\frac{\theta}{2},-\frac{2}{s_E})$, where
$s_E:= \argmax_{s>0} \frac{s a_0(\frac{2}{s})}{4}+1-sE$.

When $s \to 0$, we have the approximation;
\begin{align}
\gamma(s)\cong \sqrt{\frac{s}{2}}-\frac{s}{16}.
\end{align}
Then, $\kappa(E)$ is approximated as
\begin{align}
\kappa(E)\cong \frac{1}{8E}-\frac{1}{128 E^2}.
\end{align}

\section{Application to uncertainty relation}
Interestingly, the relation \eqref{E27} can be used for 
the uncertainty relation between 
the position and the momentum on the periodic function space $L^2_p((-\pi,\pi])$. 
In this function space, 
the uncertainty of the position is formulated as 
the uncertainty for the pair of operators $(\cos Q, \sin Q)$ as
\begin{align}
&\Delta_\varphi^2(\cos Q,\sin Q):=
\Delta_\varphi^2 \cos Q+ \Delta_\varphi^2 \sin Q \nonumber\\
=&\langle \varphi |\cos^2 Q|\varphi \rangle
+\langle \varphi |\sin^2 Q|\varphi \rangle
-\langle \varphi |\cos Q|\varphi \rangle^2
-\langle \varphi |\sin Q|\varphi \rangle^2  \nonumber\\
=&
1-\langle \varphi |\cos Q|\varphi \rangle^2
-\langle \varphi |\sin Q|\varphi \rangle^2.
\end{align}
On the other hand, the uncertainty of the momentum is given as
 $\Delta_\varphi^2 P=
 \langle \varphi |P^2 |\varphi \rangle
-\langle \varphi |P|\varphi \rangle^2$.
Thus, the uncertainty relation is formulated as
the trade-off between 
$\Delta_\varphi^2(\cos Q,\sin Q)$ and $\Delta_\varphi^2 P$.
That is, 
this trade-off can be formulated as the following minimization
\begin{align}
\min_{\varphi\in L_p^2([-\pi,\pi))}
\{ \Delta_\varphi^2(\cos Q,\sin Q)| \Delta_\varphi^2 P \le E, \|\varphi\|=1\}.
\end{align}
Since this problem has symmetry, we can restrict our function $\varphi$
to satisfy the conditions 
$\langle \varphi |\sin Q|\varphi \rangle=0$ and 
$\langle \varphi |P|\varphi \rangle=0$.
Then, our problem is simplified to 
\begin{align}
&\min_{\varphi\in L_p^2([-\pi,\pi))}
\{ 1-\langle \varphi |\cos Q|\varphi \rangle^2
| \langle \varphi |P^2 |\varphi \rangle \le E, \|\varphi\|=1\}  \nonumber\\
=&
1-\Big(\max_{\varphi\in L_p^2([-\pi,\pi))}
\{ \langle \varphi |\cos Q|\varphi \rangle
| \langle \varphi |P^2 |\varphi \rangle \le E, \|\varphi\|=1\} \Big)^2 \nonumber\\
=& 1-\kappa(E)^2.
\end{align}
By using \eqref{E27},
this trade-off is solved as the following relation \cite[Theorem 10]{CMP}.
\begin{align}
\min_{\varphi\in L_p^2([-\pi,\pi))}
\{ \Delta_\varphi^2(\cos Q,\sin Q)| \Delta_\varphi^2 P \le E, \|\varphi\|=1\}
=\max_{s>0} 1- 
\Big(sE -\frac{sa_0(2/s)}{4} 
\Big)^2.\label{MAA}
\end{align}
In addition,
the minimum in \eqref{MAA} is attained when and only when the function $\varphi$ 
is given as a shift of the Mathieu function $\ce_0(\theta 2 ,-\frac{ 2}{s_E})$. 
Moreover, the right hand side of \eqref{MAA} is asymptotically expanded as $\frac{1}{4E}-\frac{1}{32E^2}$
when $E$ goes to infinity.

\section{Conclusion}
This paper explains several applications of special functions to phase estimation.
In particular, we have addressed 
prolate spheroidal wave function and 
Mathieu function.
Although Mathieu function works for phase estimation under a certain energy constraint,
it also works for the estimation of the unknown unitary under a certain energy constraint
when the set of unknown unitaries form a group representation of SU(2) \cite{CMP}.

Another type of energy constraint for phase estimation problem 
was discussed in the reference \cite{HVK}.
This problem setting uses a function related to Gamma function.
In this way, 
special functions have various applications in quantum information.
As another example of special functions to quantum information, 
the reference \cite{HAY} studied the relation between 
Askey scheme and quantum state distinguishability.
It is expected that more special functions will be applied to the analysis on various types of quantum information processings.

All the presented results assume the noiseless case.
While the Heisenberg scaling with the noisy case was studied in \cite{HLY},
the relations with special functions were not studied in the noisy case.
Therefore, it is an open problem to extend 
these relations to the noisy case.

\begin{acknowledgement}
The author was supported in part by the National
Natural Science Foundation of China (Grants No. 62171212) and
Guangdong Provincial Key Laboratory (Grant No. 2019B121203002).
\end{acknowledgement}
%

\input{references}

\end{document}

%% file: references.tex
%%%%%%%%%%%%%%%%%%%%%%%% referenc.tex %%%%%%%%%%%%%%%%%%%%%%%%%%%%%%
% sample references
% %
% Use this file as a template for your own input.
%
%%%%%%%%%%%%%%%%%%%%%%%% Springer-Verlag %%%%%%%%%%%%%%%%%%%%%%%%%%
%
% BibTeX users please use
% \bibliographystyle{}
% \bibliography{}
%